# Collapse of the $Gd^{3+}$ ESR fine structure throughout the coherent temperature of the Gd-doped Kondo Semiconductor $CeFe_4P_{12}$

P. A. Venegas[1], F. A. Garcia[2], D. J. Garcia[3], G. Cabrera[4], M. A. Avila[5], and C. Rettori[4,5]

[1]UNESP-Universidade Estadual Paulista, Departamento de Física, Faculdade de Ciencias, C.P. 473, Bauru-SP 17033-360, Brazil.

[2] Universidade de São Paulo, IFUSP, BR-05508090 São Paulo, SP, Brazil.

[3]Centro Atómico Bariloche (CNEA) and Instituto Balseiro (U. N. Cuyo), CONICET, CP 8400 Bariloche, Río Negro, Argentina.

[4]Instituto de Física "Gleb Wataghin", UNICAMP, 13083-859, Campinas, SP, Brazil.

[5]CCNH, Universidade Federal do ABC (UFABC), 09210-580 Santo Andre, SP, Brazil.

ABSTRACT

Experiments on $Gd^{3+}$ Electron Spin Resonance (ESR) in the filled skutterudite $Ce_{1-x}Gd_xFe_4P_{12}$ ($x \approx 0.001$), at temperatures where the host resistivity manifests a smooth *insulator-metal* crossover, provides evidence of the underlying Kondo physics associated with this system. At low temperatures (below $T \approx 160$ K), $Ce_{1-x}Gd_xFe_4P_{12}$ behaves as a Kondo-insulator with a relatively large hybridization gap, and the $Gd^{3+}$ ESR spectra displays a fine structure with *lorentzian* line shape, typical of insulating media. The electronic gap is attributed to the large hybridization present in the coherent regime of a Kondo lattice, and Mean-Field calculations suggest that the electron-phonon interaction is fundamental at explaining such hybridization. The resulting electronic structure is strongly temperature dependent, and at $T^* \approx 160$ K the system undergoes an insulator-to-metal transition induced by the withdrawal of 4*f*-electrons from the Fermi volume, the system becoming metallic and non-magnetic. The $Gd^{3+}$ ESR fine structure coalesces into a single *dysonian* resonance, as in metals. Still, our simulations suggest that exchange-narrowing via the usual Korringa mechanism, is not enough to describe the thermal behavior of the $Gd^{3+}$ ESR spectra in the entire temperature region (4.2 – 300 K). We propose that temperature activated fluctuating-valence of the *Ce* ions is the key ingredient that fully describes this unique temperature dependence of the $Gd^{3+}$ ESR fine structure.

**Key words:** *Fluctuating valence; narrowing effects; Skutterudites; Kondo insulator; insulator-to-metal transition; activated relaxation.*

PACS number(s): 71.27. + a, 75.20.Hr, 76.30. − v

INTRODUCTION

Among the strongly correlated electron systems, several rare-earth compounds, known as hybridization gap semiconductors or Kondo-insulators/semiconductors, have recently attracted great interest [1–11]. Kondo-insulators belong to a class of strongly correlated materials that form a group of either nonmagnetic semiconductors with a *narrow-gap* or semimetals with tiny gaps. At low-$T$, the Kondo effect develops coherence throughout the system, forming a renormalized Fermi liquid, where conduction and 4*f*-electrons contribute to the counting for the volume of the Fermi surface. Differently from heavy fermions systems, in Kondo-insulators the heavy-electron band is completely filled and the chemical potential falls in the middle of the hybridization gap (half-filling condition [12]). This semiconducting state manifests when the pseudogap straddles the Fermi energy, and is subjected to many-body renormalizations, leading to a *T*-dependent reduction of its magnitude [1,3–11,13–17]. Among new and interesting compounds with properties being related to this model are Ce-based filled skutterudites compounds [18]. They have the chemical formula $RM_4X_{12}$ and crystallize in the $LaFe_4P_{12}$ structure, with space group $Im\bar{3}$ and local point symmetry $T_h$ for the rare-earth (R) ions [19]. The R cations are usually referred to as *guest* or *filler* ions, and reside within the voids of the [$M_4X_{12}$] polyanion *host* framework, or *cage* structure. Based upon simple electron counting, one can conclude that the series of Ce-based filled skutterudites $CeM_4X_{12}$ for which M = Fe, Ru, Os and X = P, As, Sb, are systems presenting a half-filled band in the presence of a 4*f*-electron.

The low-$T$ transport properties of these materials are similar to conventional semiconductors, but in contrast to what is expected from simple thermal activation, the gap seems to disappear at a temperature $T^*$, which is low relative to the gap size, indicating strong collective behavior [20,21]. Concerning $CeFe_4P_{12}$, its semiconducting properties at low-$T$ were considered anomalous when compared to other isostructural members of this class of compounds, most of which are metallic and magnetic [22]. The resistivity varies over six orders of magnitude between 50 K and 300 K, but the data can only be fitted to an activated conduction process over a limited temperature range. This fitting yields a pseudogap of about 1500 K in magnitude, roughly three times smaller than the value predicted by LDA calculations [23], but in agreement with other experiments [24]. The magnetic susceptibility is nearly $T$-independent over the range $100\,K < T < 300\,K$, clearly indicating that the ground state is non-magnetic.

Lattice parameters deviations indicate mixed-valence character for the 4$f$-element in all cases, which has also led to the definition of intermediate-valence semiconductors [25]. Recent studies of *Ce* and *Sm* compounds in this class of materials led to a renewed interest in this old hybridization gap (pseudogap) problem [1,2] and also to new, exotic types of Kondo effects [26]. Photoemission Spectroscopy on single crystals of $CeFe_4P_{12}$ [1,22] showed a strong mixed-valence behavior of Ce ions, with a three-peak structure for the 4$f$-level, with the presence of $4f^0, 4f^1$, and $4f^2$ configurations due to the strong hybridization with the band. They found a mean occupation $n_f = 0.86$ for the 4$f$-level, which is near the trivalent case as the nominal valence (value of 3.14). In addition, the anomaly observed in the lattice parameter [3,27] and results from band structure calculations [23] are also compatible with an intermediate-valence for the Ce ions.

A topic of growing interest in the field of the skutterudites is to understand the interplay between their lattice dynamics and physical properties. Local vibrations of R cations, presumably indicate a strong electron-phonon coupling that should be taken into account in any realistic picture. In $Ce_{1-x}Yb_xFe_4P_{12}$, a rattling behavior is believed to account for the 20% reduction of the hyperfine parameters for the two $Yb^{3+}$ sites observed in the electron spin resonance (ESR) experiments [28]. As for $La_{1-x}Gd_xPt_4Ge_{12}$, it was suggested to explain the evolution of the ESR linewidth as a function of temperature [29]. In any case, the energy scale of these vibrations are close to the so-called Einstein temperature ($\theta_E$), obtained through the independent rattler approximation. In the case of *Ce*-based skutterudites, this characteristic energy scale is about 100 K and, in the particular case of $CeFe_4P_{12}$, $\theta_E = 148$ K, as probed by EXAFS studies [30]. Surprisingly, this temperature scale is about the same as the temperature $T^* \approx 150$ K, which marks the loss of coherence of heavy-fermion behavior, obtained from resistance experiments [20].

In a previous work on the ESR of $Gd^{3+}$ in $Ce_{1-x}Gd_xFe_4P_{12}$ [31], it was shown that the low-$T$ $Gd^{3+}$ ESR spectra, composed of seven resolved insulator-like *lorentzian* lines, present temperature-activated behavior that yields a striking coalescence of the resolved fine structure into a single *dysonian* (metallic-like) resonance at $T^* \approx$ 150-160 K. To our knowledge, it was the first report of such a coalescence and change in the ESR lineshape at high-$T$. This behavior is certainly related to the insulator-to-metal crossover induced by the loss of coherence, and exhibits the sensitivity of EPR experiments to probe such a remarkable effect. In this paper, we present numerical simulations for the collapse of the crystal field fine structure of the $Gd^{3+}$ ESR spectra in $Ce_{1-x}Gd_xFe_4P_{12}$ ($x \approx 0.001$) in the temperature range of interest. Our simulation shows the subtle and concomitant interplay between the *Ce* 4*f* fluctuation-valence (FV) and exchange-narrowing (EN) effects. The loss of coherence of the Fermi liquid state, yielding a smooth insulator-metal transition at $T^* \approx 160$ K, gives further support to the assignment of $CeFe_4P_{12}$ as a Kondo-semiconductor with a relatively large and *T*-dependent *pseudogap* [1,30,32]. This *T*-dependence originates from many-body correlation effects in a Kondo lattice. At low-*T*, long-range coherence effects develop among *f*-electrons, so that they participate in band properties and a gap opens due to hybridization with the conduction band. At half-filling, the valence band is completely filled if one counts *f* and conduction electrons (*ce*) together, and the system is an insulator. When the temperature is increased, changes of the electronic structure are produced, with the progressive filling of the gap at a lower temperature than would be expected from the simple thermal activation picture in conventional semiconductors [21]. An insulator-to-metal transition is induced at a coherence temperature $T^*$ much lower than the Kondo temperature, marking the drop out of *f*-electrons from the Fermi volume, leaving behind conducting holes in the valence band and localized *f*-electrons at the R ions [33]. In this regime, *f*-electrons are nearly decoupled from the conduction band and stay in localized states relatively away from the Fermi level, which is purely of *ce*-nature, with a nonzero, but small density of states at $E_F$. We illustrate the above scenario with a hypothetical density of states (DOS) in Fig. 1, displaying the cases of temperatures below and above $T^*$. Details of the calculations are presented in [34]. The role of lattice dynamics in affecting the electronic properties through strong coupling effects will be discussed later on.

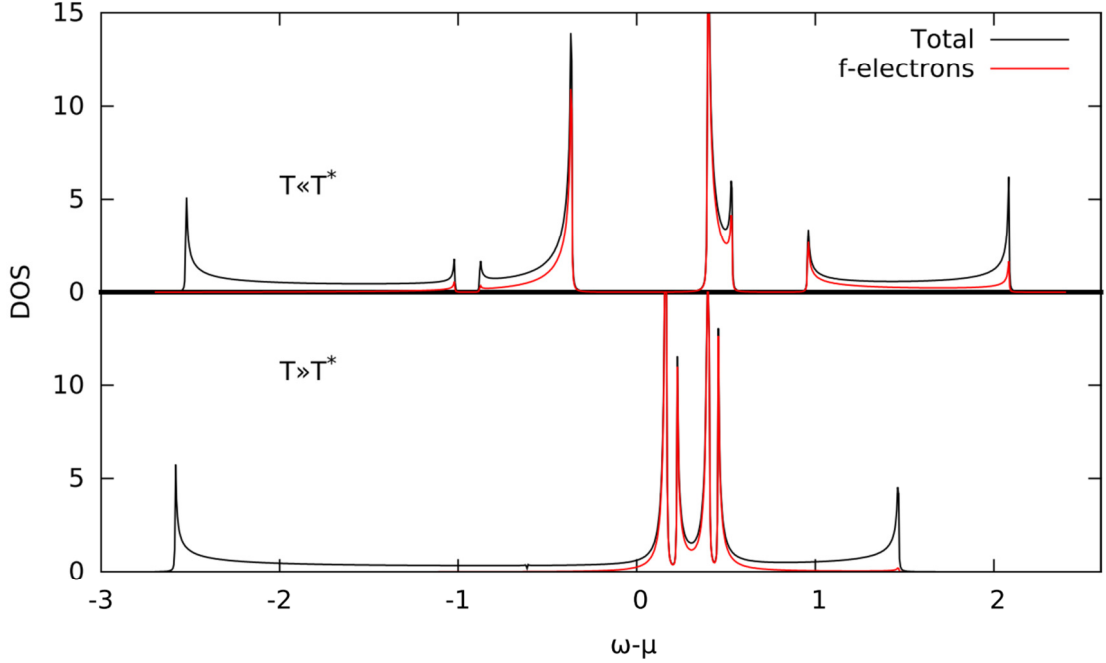

FIG. 1 (Color online). Temperature dependent properties of the DOS, showing in red the partial contribution of 4*f*-electrons. For the low-*T* state ($T \leq T^*$), Ce 4*f*-electrons participate in band properties. Hybridization effects are enhanced by the electron-phonon interaction, and a gap opens in the DOS (higher panel), the system becoming a Kondo insulator. The lower panel shows the high-*T* ($T \geq T^*$) behavior, where Ce 4*f*-electron states decouple from the band and become localized, with a vanishingly small hybridization. The chemical potential µ moves inside the valence band, while the gap closes. Details are presented in [34].

RESULTS AND ANALYSIS

The analysis of the ESR spectra of $Gd^{3+}$ diluted in $CeFe_4P_{12}$ from 4.2 K to 300 K requires to consider three temperature regions to understand the *T*-dependence of the observed spectra: *i)* below 150 K, the ESR spectra is consistent with $Gd^{3+}$ ions diluted in insulators, *i.e.,* there is no relaxation via *ce* (Korringa relaxation) [35,36]. The line shape associated to each fine structure of the spectrum is *lorentzian* and the full resolved spectra can be accounted for by the well known anisotropic spin Hamiltonian for cubic symmetry [37]:

$$\mathcal{H} = \mu_B HgS + \frac{1}{60} b_4 \left( O_4^0 + 5 O_4^4 \right) \tag{1}$$

where the first term is the Zeeman energy, the second term is the fourth-order contribution of the cubic crystal field, with $b_4$ the fourth-order crystalline field parameter for $Gd^{3+}$, and $O_4^0$ and $O_4^4$ the spin operators of fourth degree; *ii)* between ~150 and ~200 K, the observed fine structure coalesces into a single line of *dysonian* shape, characteristic of ESR of diluted magnetic moments in metals [38,39]; *iii)* above ~200 K the single *dysonian* line presents a nearly linear thermal broadening of the linewidth of ~ 1 Oe/K, characteristic of a Korringa-like relaxation via conduction electrons.

Since the system is undergoing a 'smooth' insulator-metal transition at about 150 K, the exchange interaction between $Gd^{3+}$ local moments and *ce* must be considered. The Plefka-Barnes (PB) model, for the EN of the fine structure, is normally attempted to simulate such spectra [40–42]. Although we found that the PB model gives a good description of the thermal broadening of the linewidth for $T \geq 200$ K, it is not capable of reproducing the experimental details for the coalescence of the fine structure observed in the ESR spectra between 150 K and 200 K [31]. Two discrepancies between the calculated and experimental spectra between 150 K and 182 K can be noted: *i)* using the PB theory the spectra still show a partial resolved fine structure where the experiment already displays a collapsed spectrum; *ii)* the experiment shows a nonlinear thermal broadening of the spectra at the angle of collapsed fine structure (30º from the [001] direction in the (110) plane, see Fig. 2), whereas the PB theory shows a linear increase. Therefore, these results suggest that, besides the crystal field and exchange coupling effects, one needs an additional and complementary mechanism in order to understand the coalescence of the $Gd^{3+}$ ESR fine structure between ~ 150 K and ~ 200 K.

The missing ingredient that can satisfactorily explain the experimental results is the fluctuating valence (FV) of the *Ce* ions in the filled skutterudite $CeFe_4P_{12}$. The mechanism leading to this FV is usually ascribed to the formation of the hybridization gap. However, in skutterudites, there is also the intriguing possibility that the coupling between the electronic state and the low energy vibrations of the lattice may lead to FV and Kondo-like phenomena. In $CeFe_4P_{12,}$ the picture obtained from spectroscopic experiments [1,22], indicates a large *f-ce* hybridization strength. Below $T^*$, lattice vibrations cooperate coherently with the electronic insulating state, favoring the opening of the hybridization gap. Due to the large hybridization, Ce *4f*-electrons participate in band properties, hopping on and off between *f* states and the band, with strong fluctuations of the Ce valence. Despite one of the configurations being magnetic, it is known that these strong fluctuation effects quench the magnetic susceptibility [17,43], which remains practically constant with temperature, as observed in experiments [20]. This picture is also consistent with the unusual transport properties obtained in [44] for the thermoelectric power and the Hall carrier mobility, which are attributed to the 4*f*-electron hybridization.

When $T \gtrsim T^*$, our model indicates that *f*-states decouple from the band and stay above the Fermi energy, the system being metallic (with a small density of states at $\mu$) and non-magnetic. Rattling of *Ce* ions is now incoherent and 4*f*-states become localized. Since the hybridization is now small, the FV now implies an activation energy to move a band electron from the Fermi level to the *f*-states (see Fig. 1). Calculations shown in [34] indicate that the above activation energy is weakly *T*-dependent and can be approximated by its value at the insulator-to-metal transition (which we call $E_{ex}$).

Therefore, there are enough indicia to include in the simulation the effect of the FV of Ce ions on the $Gd^{3+}$ ESR spectra. The formalism developed by Venegas *et al.* [45], can be considered as an additional *T*-dependent relaxation mechanism caused by the exchange interaction between the $Gd^{3+}$ ion and the fluctuating Ce magnetic moments. This contribution to the $Gd^{3+}$ ESR linewidth is approximated by an activation law, and is written as:

$$\Delta H_{FV} = A e^{-E_{ex}/kT}, \qquad (2)$$

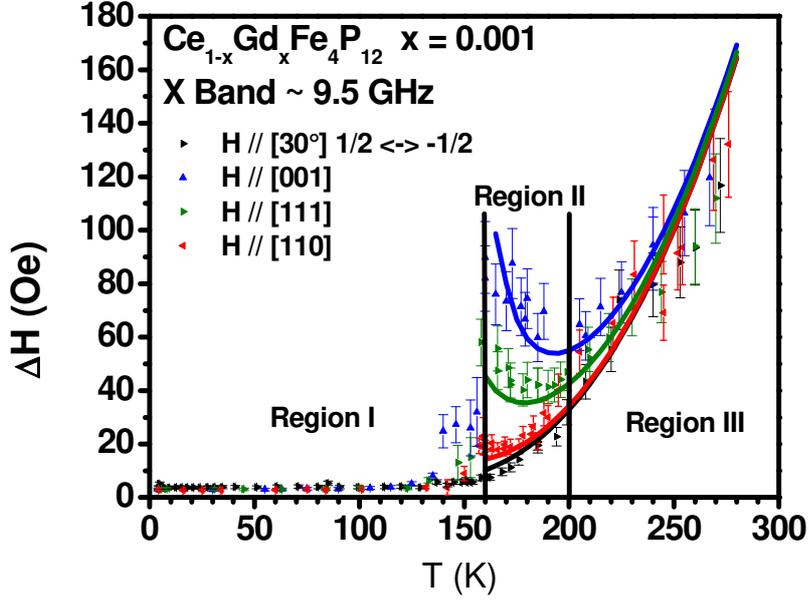

FIG. 2 (Color online). ESR linewidth of $Ce_{1-x}Gd_xFe_4P_{12}$ ($x \approx 0.001$). Region I for the collapsed ESR fine structure ($H//[30º]$) and transition $-½ \leftrightarrow ½$ for the other main crystal orientations. Overall full linewidths from the theoretical simulations in Regions II and III.

where $A$ is a constant and $E_{ex}$ is the interconfigurational excitation energy discussed above.

Results for $\Delta H$ are shown in Fig. 2, for several orientations of the field. At low-$T$, in the insulating regime, $T \leq T^*$ (Region I), the linewidth of the $-½ \leftrightarrow ½$ transition is nearly constant, equal to a residual value, $\Delta H_{res}$. At $T \approx T^*$, contributions including the Korringa rate, exchange narrowing (EN) effects, and activated FV mechanism begin to appear. The crystal field effects on the $Gd^{3+}$ fine structure are responsible for the angular dependence of $\Delta H$, and at about $T \approx 200$ K, the line becomes nearly isotropic. An *apparent* Korringa behavior is displayed for $T \geq 200$ K, but our simulation suggests that the increase of the linewidth in Region III should be ascribed to the activated FV exchange. The calculation of the overall ESR linewidth between ~ 150 K and ~ 200 K, shown in Fig. 2, is much closer to the experimental data and considerably improved in comparison with the data analysis shown in Ref. [31].

The simulated spectra, within the same temperature range, are presented in Fig. 3. The agreement with experimental results from Ref. [31] is remarkable, including the collapse of the individual $Gd^{3+}$ ESR fine structure lines due to both EN and FV processes.

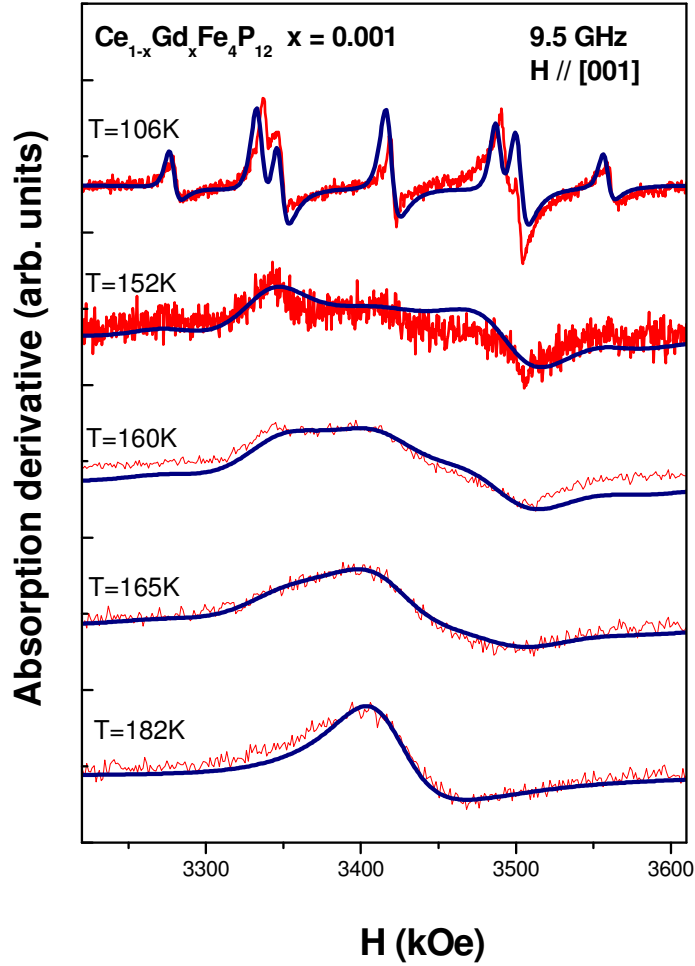

FIG. 3 (Color online). ESR spectra of $Ce_{1-x}Gd_xFe_4P_{12}$ ($x \approx 0.001$). Full lines represent the theoretical model simulations.

We have assumed that the conduction band starts becoming appreciably populated at $T^* = 160$ K. Therefore, the Korringa contribution to the linewidth starts at $T^*$ and increases linearly with temperature, as $b(T - T^*)$. Following Urban *et al.* [31], narrowing effects are included in a phenomenological way, mimicking an effective Gd-Gd interaction, with a mean exchange field of 0.035 Oe and a mean width of ~ 100 Oe, for the distribution of exchange fields. The overall linewidth is fitted to the expression:

$$\Delta H = \Delta H_{res} + b(T - T^*) + A e^{-\frac{E_{ex}}{kT}}, \quad (3)$$

where $\Delta H_{res}$ is the residual linewidth, $b$ is the Korringa parameter, $A$ is a constant and $E_{ex}$ is the interconfigurational excitation energy discussed above. The main parameters for optimal fitting obtained from simulations are listed in Table I:

| $b$ [Oe/K] | $\Delta H_{res}$ [Oe] | $E_{ex}$ [K] | $A$ [Oe] | $b_4$ [Oe] | $g$ |
|---|---|---|---|---|---|
| 0.05 | 5 | 1180 | 11000 | 7 | 1.987 |

Table I: Parameters obtained from the simulation of experimental results of the ESR spectra of Gd diluted in $CeFe_4P_{12}$. Values were extracted from calculations in the temperature range $4\,K \leq T \leq 300$ K. The table includes parameters from equations (1) and (3).

From Table I, the extracted Korringa-rate is much smaller than the *apparent* Korringa-rate of ~1 Oe/K obtained for $T \gtrsim 200\,K$. This is of key importance. The Korringa-rate depends on $J_{fs}\eta_F$, the product of the *f-ce* exchange interaction, and the density of states at the Fermi level. Then, the small extracted value implies in either small $J_{fs}$, small $\eta_F$, or both. Therefore, the relatively large variation of $\Delta H$ for $T \gtrsim 200\,K$ is mainly associated with the influence of the FV mechanism of Ce ions on the $Gd^{3+}$ ESR spectra. The extracted interconfigurational excitation energy of the Ce ions, $E_{ex}$ = 1180 K, illustrates the nature of the pseudogap as a full many-body concept. In the incoherent regime of the Kondo physics, *f* and *ce* states are mostly decoupled. While the Fermi surface is of *ce* nature, with a small density of states at the chemical potential µ, the excitation energy $E_{ex}$ obtained in the simulation measures the distance from the Fermi level to the isolated *f*-electronic states. Remarkably, the ESR experiment captures both aspects of the same phenomenon: Korringa rate and EN effects due to *ce*, and FV exchange contributions (Eq. 2) which come from the physics of the *f*-electronic structure.

CONCLUSIONS

Based on our ESR experiments in the $Ce_{1-x}Gd_xFe_4P_{12}$ ($x \approx 0.001$) skutterudite, we propose that the hybridization gap is indeed strongly *T*-dependent and closes with increasing temperatures. At $T = T^* \approx 160$ K, the gap is not completely closed, but the system undergoes an insulator-to-metal transition due to the drop out of *f*-electrons from the Fermi volume, with the chemical potential shifting inward the valence band (see Fig. 1). This temperature $T^*$ marks the loss of coherence, when *f*-states decouple from the band and localize above the Fermi level. As a consequence, the system remains non-magnetic due to the small weight of *f*-states, which are practically empty in the ground configuration. Since the hybridization is small for $T \geq T^*$, the high energy scale $E_{ex}$ being probed by ESR experiments (of the order of 1000 K), is to be associated with the activated promotion of a *ce* from the Fermi level to the $4f^1$ magnetic configuration, leaving a hole in the conduction band. The presence of a temperature activated FV effect renders an extra exponential thermal broadening for the $Gd^{3+}$ ESR linewidth. This allows us to simulate the coalescence of the $Gd^{3+}$ fine structure in the ESR spectra and the change of the resonance lineshape from *lorentzian* (insulator-media) to *dysonian* (metallic-media) at about the same temperatures where a smooth *insulator-metal* transition takes place. Just as importantly, by means of ESR measurements, we have obtained for the first time unequivocal experimental evidence for the presence of strong FV in a Kondo-semiconductor, with the simultaneous identification of the coherence temperature $T*$ of the Kondo lattice.

ACKNOWLEDGEMENTS

The authors thank FAPESP (SP-Brazil), CNPq and CAPES (Brazil), for financial support and to NCC/GridUnesp for the computing resources.

# Supplemental Material for: Collapse of the $Gd^{3+}$ ESR fine structure throughout the coherent temperature of the Gd-doped Kondo Semiconductor $CeFe_4P_{12}$

P. A. Venegas[1], F. A. Garcia[2], D. J. Garcia[3], G. Cabrera[4], M. A. Avila[3], and C. Rettori[4,5]

[1]UNESP-Universidade Estadual Paulista, Departamento de Fısica, Faculdade de Ciencias, C.P. 473, Bauru-SP 17033-360, Brazil.

[2] Universidade de Sao Paulo, IFUSP, BR-05508090 Sao Paulo, SP, Brazil.

[3]Centro Atómico Bariloche (CNEA) and Instituto Balseiro (U. N. Cuyo), CONICET, CP 8400 Bariloche, Río Negro, Argentina.

[4]Instituto de Física "Gleb Wataghin", UNICAMP, 13083-859, Campinas, SP, Brazil.

[5]Universidade Federal do ABC, Centro de Ciencias Naturais e Humanas, 09210-170 Santo Andre, SP, Brazil.

**INTRODUCTION**

In this Supplemental Material we show a model that account for the insulator-metal transition induced by electron-phonon coupling on $CeFe_4P_{12}$. To this end we consider an array of rare earth RE *4f* orbitals (corresponding to $Ce^{3+}$ ions) coupled to a bath of conduction electrons (*ce*). To qualitatively explain the experimental results we show that a lattice distortion modifying the *4f-ce* hopping is crucial.

The simplest model for the interplay between *4f* orbitals and conduction electrons is the Periodic Anderson Model (PAM). The PAM [46,47] (also similar are the cooper-oxide like [48] and Kondo lattice models [49]) has been extensively studied. These studies show that anti-ferromagnetic or paramagnetic ground states are the probable states when the electronic concentration corresponds to half filling (one electron per orbital).

Electron-lattice coupling, the other essential ingredient, has been studied generically in the Su-Schiefer-Heeger [50,51] model. In this model lattice distortions are stable up-to some critical temperature. The particular case of the Periodic Anderson Model (PAM) where lattice distortion couple to on-site energy and hopping was discussed in Ref. [52]. That work shows that the electron-lattice coupling suppress the magnetic state.

Usual modelizations of PAM consider fully occupied RE sites with *4f* level below the Fermi level. Spectroscopic experiments [1,22] on $CeFe_4P_{12}$ found a mean occupation for the RE of $n_f = 0.86$ for the *4f* level. These experiments suggest fluctuations between $4f^0$ and $4f^1$ implying a *4f* level above the Fermi level.

In this Supplemental Materials we show that electron-lattice coupling leads to a low temperature paramagnetic insulator and a high temperature metal at half filling. To illustrate the mechanism we consider a simple one dimensional geometry although similar results can be obtained in more dimensions. The model qualitatively explains the transport and EPR properties of the *Gd* doped compound $CeFe_4P_{12}$.

**MODEL HAMILTONIAN**

To model the compound we consider a bath of conduction electrons (*ce*) coupled to RE sites. The coupling between RE and *ce* sites is modified by a lattice distortion *x*. The Hamiltonian considered reads

$$H = \sum_{\langle\langle i,l\rangle\rangle,\sigma} t_s c^\dagger_{i,\sigma} c_{l,\sigma} + \sum_{j,\sigma} E_f n_{d,j,\sigma} + \sum_{\langle\langle i,j\rangle\rangle,\sigma} t_{sf} c^\dagger_{i,\sigma} d_{j,\sigma}$$
$$+ U \sum_j n_{d,\downarrow,j} n_{d,\uparrow,j} + \lambda \sum_j L_j \cdot S_j + \sum_j 1/2 C x_j^2 + h.c.$$

where *c* operators refers to *ce* and *d* to *4f* sites orbitals. The first sum runs over the lattice sites *i* and *l* of the *ce*. The third sum runs over the lattice sites *i* of the *ce* and nearest neighbors *j* on RE of the so defined zig-zag chain. $E_f$ is the RE on-site energy, $t_s$ is the hopping between *ce* sites, and *U* is the on-site Coulomb repulsion for RE sites. $\lambda$ is the coupling between the electronic spin **S** and the orbital angular moment **L** on the RE. The distortion *x* modifies the *ce*-RE hopping as $t_{sf} = t_{sf,0} - Cg_2 x$. Assuming a uniform distortion and minimizing respect to *x* results in $t_{sf} = t_{sf,0} - g_2^2 (\langle c^\dagger d + d^\dagger c \rangle)$. Within this approximation the temperature dependence of *x* and $t_{sf}$ is entirely given by the electronic thermodynamic.

We solve the resulting electronic Hamiltonian using a mean field approximation:

$$H = \sum_{\langle\langle i,j\rangle\rangle,\sigma} t_s c^\dagger_{i,\sigma} c_{j,\sigma} + \sum_{i,\sigma} E_f n_{d,i,\sigma} + \sum_{\langle\langle i,j\rangle\rangle,\sigma} t_{sf} c^\dagger_{i,\sigma} d_{j,\sigma}$$
$$+ U \sum_i \left( n_{d,\downarrow,i} \langle n_{d,\uparrow,i}\rangle + n_{d,\uparrow,i} \langle n_{d,\downarrow,i}\rangle - \langle n_{d,\downarrow,i}\rangle \langle n_{d,\uparrow,i}\rangle \right)$$
$$+ \lambda \sum_i \left( \langle L_{z,i}\rangle \left(\frac{1}{2}(n_{d,\uparrow,i} - n_{d,\downarrow,i})\right) + h.c. \right.$$

where we decouple the on-site Coulomb repulsion as

$$n_{d,\uparrow,i} n_{d,\downarrow,i} \sim \left( n_{d,\downarrow,i} \langle n_{d,\uparrow,i}\rangle + n_{d,\uparrow,i} \langle n_{d,\downarrow,i}\rangle - \langle n_{d,\downarrow,i}\rangle \langle n_{d,\uparrow,i}\rangle \right)$$

The angular moment coupling is $L \cdot S$ (modulus of $L$ is 3 for a rare earth). We simplify this terms as $L \cdot S \sim \langle L_z \rangle S_z = \langle L_z \rangle \frac{1}{2}(n_{d,\uparrow,i} - n_{d,\downarrow,i})$. When the RE occupation is low, $\langle L_z \rangle$ should be proportionally small, so we assume $\langle L_z \rangle \propto L \langle n_{d,i} \rangle$ ($n_{d,i} = n_{d,\uparrow,i} + n_{d,\downarrow,i}$). To allow for a possible anti-ferromagnetic order we consider two different sites for *ce* and RE (a four site basis). To this end we associate alternating angular moments $\pm \langle L_z \rangle$ to the RE sites (The inclusion of the spin-orbit coupling is not essential. The main result is reproduced if this term is excluded: The insulator-metal transition with the lattice coupling is independent of $\lambda$. Without this $L \cdot S$ coupling the AF ordering is lost for the small value of $U$ consider here. Any small value induces in the uncoupled case an AF ground state as expected from exact diagonalization methods [46]). We take a global density of one electron per site (4 electrons on the four sites basis). The simplified Hamiltonian decouples up and down electrons. Due to the proposed alternating angular moments the Hamiltonian has an up/down symmetry between different *f*-sites:

$$n_{d,\sigma,1} = n_{d,-\sigma,2} \text{ where } \sigma = \uparrow \text{ or } \downarrow.$$

To illustrate the physics of the Insulator-Metal transition on CeFe$_4$P$_{12}$ we use a zig-zag chain (see Fig. S1).

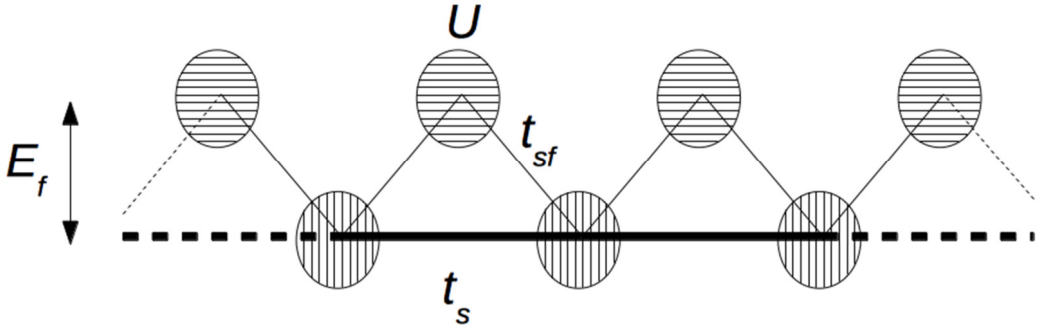

FIG. S1. Circles indicate *ce* (bottom) and RE (top) sites. Parameters are shown on the Figure.

The Hamiltonian can be readily diagonalized on Fourier space. The resulting *k*-space Hamiltonian is

$$H = \sum_{k,\sigma} E_{f,\sigma,1} n_{d1,\sigma,k} + E_{f,\sigma,2} n_{d2,\sigma,k} + t_s \left(1 + e^{Ik}\right) c^\dagger_{1,\sigma,k} c_{2,\sigma,k}$$
$$+ t_{sf} \left( c^\dagger_{1,\sigma,k} d_{1,\sigma,k} + d^\dagger_{1,\sigma,k} c_{2,\sigma,k} + c^\dagger_{2\sigma,k} d_{2\sigma,k} + e^{Ik} d^\dagger_{2,\sigma,k} c_{1,\sigma,k} \right) + h.c.$$

where $c^\dagger_{1,\sigma,k}(d_{1,\sigma,k})$, $i=1,2$, correspond to the destruction operators on the two different *ce* (RE) sites of the basis. We take $t_s = 1$ as our energy unit. The parameters $t_{sf,0} = 0.05$ and $E_f = 0.5$ are chosen as to leave almost empty the *4f* orbitals without any further interaction and be compatible with a *4f* state fluctuating between $4f^0$ and $4f^1$. The parameters $\frac{1}{2}\lambda L = 0.1$ and $U = 0.5$ are of order of magnitude of the size of the bandwidth as expected for RE ions [53]. $g_2$ takes the values 0 in the uncoupled case or $g_2 = 1$ for the strongly coupled situation. In this mean field approximation the on-site energies of the *4f* orbitals are $E_{f,\sigma,1} = E_f + \frac{1}{2}\lambda L \langle n_{d1}\rangle + U\langle n_{d,-\sigma,1}\rangle$ and $E_{f,\sigma,2} = E_f - \frac{1}{2}\lambda L \langle n_{d2}\rangle + U\langle n_{d,-\sigma,2}\rangle$. $E_{f1}$ and $E_{f2}$ must be determined self-consistently.

**RESULTS**

Figure S2 shows the low and high temperature dispersion relation with ($g_2 = 1$, dotted line) and without ($g_2 = 0$, full line) electron-lattice coupling. Flatter bands have mostly *4f* character. The two flat bands split due to the spin orbit coupling and the Coulomb repulsion. For $g_2 = 1$, bands have a mixed character with a large participation of *4f* electrons (see also Fig. 1 of the main text). This last state is known as the coherent regime in the Kondo and heavy fermions systems.

The electron-lattice coupling greatly enhances the hybridization and induces a gap between the valence and conduction bands at low *T*. With one particle per site the chemical potential falls between the second and third band (zero value in the energy axis). At this temperature the coupled system is an insulator. The gap between the second and third band is roughly proportional to $t_{sf}^2$. Without electron-lattice coupling the low temperature gap is proportional to $t_{sf,0}$. The much larger gap in the coupled case is due to the fact that $t_{sf}$ is dominated by $g_2^2\left(\langle c^\dagger d\rangle + h.c.\right)$, which in turn is much larger than $t_{sf,0}$.

At high temperature both coupled and uncoupled dispersions are similar. This is due to a vanishing hybridization $\langle c^\dagger d\rangle$ with temperature (Fig. S3a), effectively decoupling electrons and lattice. The $g_2 = 1$ case becomes metallic as the highest state of the valence band (*f* on Figure S3b) moves above the Fermi level ($T_c(g_2 = 1) \sim 0.4$). The uncoupled case is always metallic as the chemical potential falls inside the third band at low temperatures ($T < 0.4$) and inside the second band at higher temperatures.

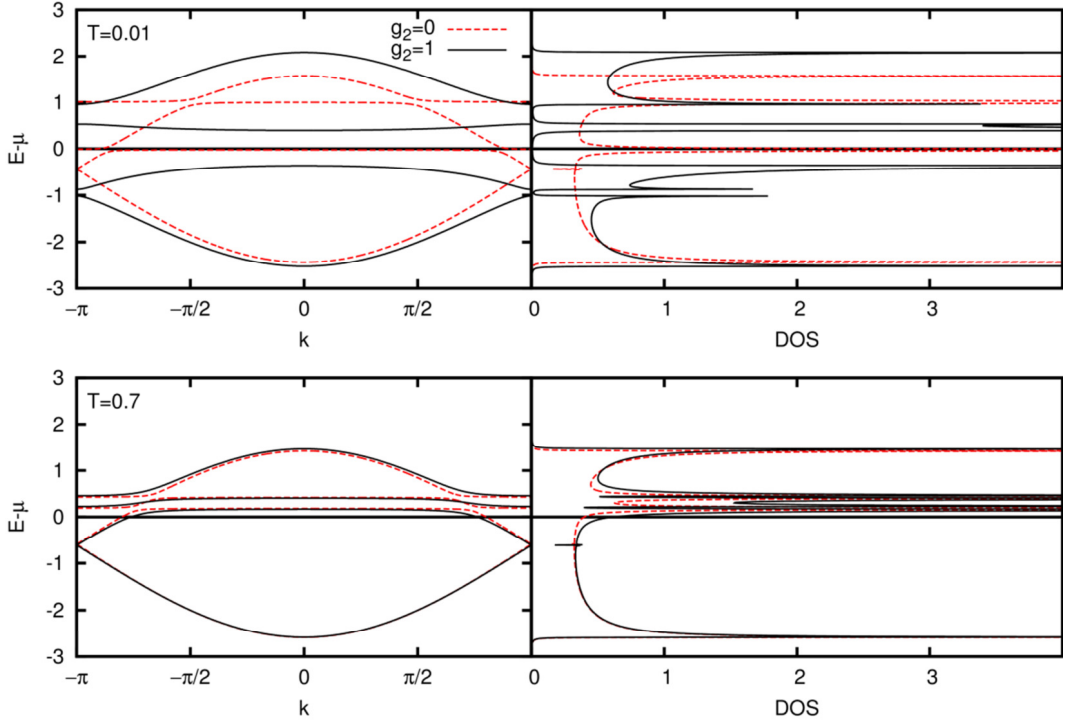

FIG. S2 (Color online). Dispersion relation (left panels) and corresponding Density of States (DOS $= \frac{-1}{\pi} \text{Im}[G(\omega + 0.002I)]$, right panel). Full black (dashed red) lines correspond to the coupled (uncoupled) lattice and electron case. Top panels correspond to $T = 0.01$. Low panels correspond to $T = 0.7$.

The net charge on the *ce* and RE sites is similar in the coupled and uncoupled cases (Fig. S3c). The RE charge is close to 0.8. With $g_2 = 0$ the RE magnetic moment is almost completely polarized on each site (Fig. S3d). This magnetic moment decreases abruptly at high temperatures signaling an anti-ferromagnetic transition ($T_N(g_2 = 0) \sim 0.3$). When $g_2 = 1$ the RE polarization is strongly reduced. The temperature dependence of the RE polarization is weaker, with no appreciable variation at the insulator-metal transition temperature $T_c$. We identify the $g_2 = 1$ state with a paramagnetic state.

In the coupled case the gap follows the *ce*-RE hybridization. Both gap and hybridization decrease as the temperature increases. The hybridization goes to zero smoothly around $T \sim 0.4$ (coherence temperature) in the $g_2 = 1$ case, while in the uncoupled case the hybridization remains small. The gap between valence and conduction bands (Fig. S3b) is almost zero for the uncoupled case, while it has a strong temperature dependency for $g_2 = 1$. The gap never closes for $g_2 = 1$.

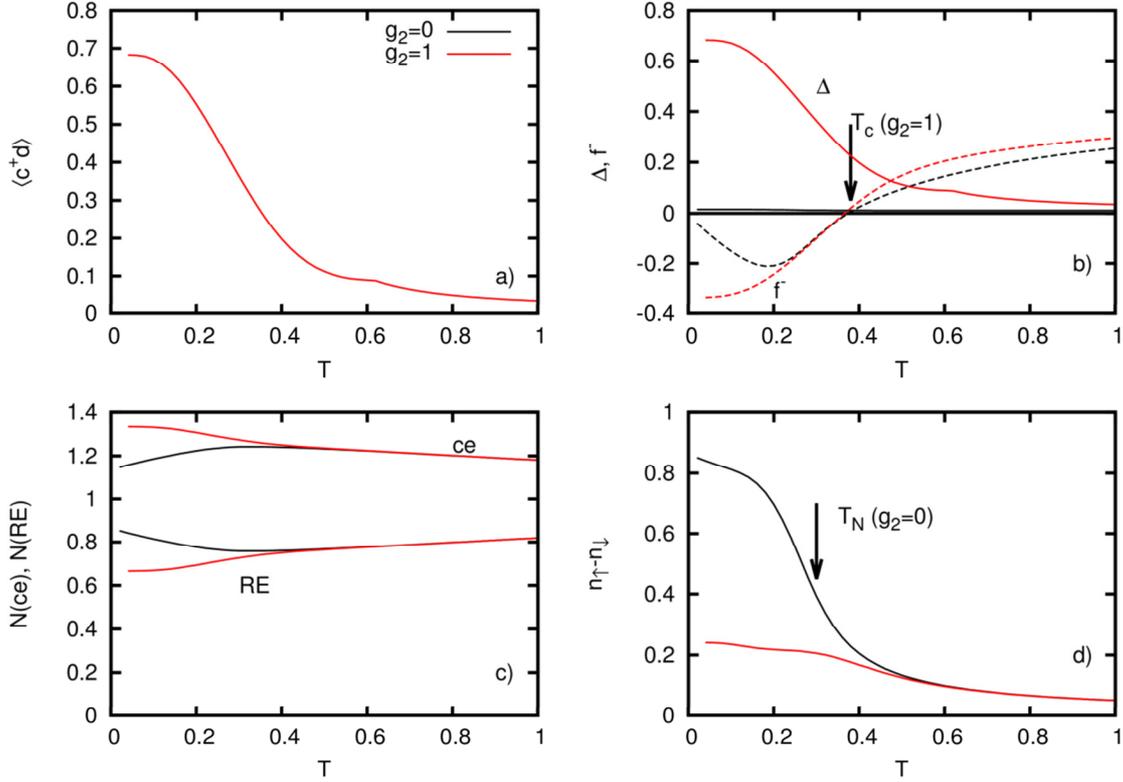

FIG. S3 (Color online). Effect of temperature $T$ on various properties of the model: a) Hybridization. b) gap (continuous line) and highest energy of the second band (dashed line, $f$) respect to the Fermi level. c) electron density on a conduction electron (upper curves) and RE (lower curves) sites. d) Polarization $n_{d1,\uparrow} - n_{d1,\downarrow}$. Black (red) line corresponds to the coupled (uncoupled) case. Hybridization on panel a) and charge gap on panel b) for the uncoupled case are close to the zero axis.

**DISCUSSION**

A paramagnetic or AF ground state is expected from numerical results on the Anderson Model [47] and Kondo Lattice Model [49]. Our results show that an Anderson-like [46] (or more precisely copper-oxide-like [48]) model coupled to lattice distortions describe an insulator-metal (I-M) transition driven by the shift of the chemical potential. In agreement with Ref. [52], our results show that the magnetic order (anti-ferromagnetic) is suppressed with the electron-lattice coupling. As expected [51,52] our solution shows a long range distortion at low *T*. The larger variation of the *ce*-RE hybridization with *T* when $g_2 \neq 0$ is consistent with numerical results of Ref. [47].

We have made no attempt to use realistic values for the parameters, as only a qualitative argumentation was presented. Nevertheless, the present approach clarifies a key point, showing that the conduction electron-lattice coupling is essential to have an insulator-metal transition in a paramagnetic state. The insulating state with a large charge gap at low *T* explains the almost temperature independent line width of the EPR experiments in this temperature range. The I-M transition marks the onset of a Korringa broadening at high temperatures. The activated behavior of valence fluctuations on RE (*ce* to RE) in the metallic regime is explain as the *4f* bands do not form a wide band, but instead well localized states above the Fermi level.

**CONCLUSIONS**

The insulator to metal transition reported on the main text is qualitatively explained as the result of the coupling between the rare earth ions – conduction electrons hopping and the lattice distortion. At the transition temperature the gap between bands does not close. Instead the Fermi level falls below the largest valence state resulting in a metallic state. This explains the exponential fluctuations on the rare earth valence and global metallic behavior of $CeFe_4P_{12}$ above the insulator-metal transition as seen by EPR.